\documentclass[a4paper]{spie}  

\usepackage{amsmath,amsfonts,amssymb}
\usepackage{graphicx}
\usepackage[colorlinks=true, allcolors=blue]{hyperref}

\title{Improving segmentation of calcified and non-calcified\\plaques on CCTA-CPR scans via masking of the artery wall}

\author[a,b]{Antonio Tejero-de-Pablos}
\author[b,a]{Hiroaki Yamane}
\author[a,b]{Yusuke Kurose}
\author[c]{Junichi Iho}
\author[c]{Youji Tokunaga}
\author[c]{Makoto Horie}
\author[c]{Keisuke Nishizawa}
\author[c]{Yusaku Hayashi}
\author[c,b]{Yasushi Koyama}
\author[a,b]{Tatsuya Harada}
\affil[a]{The University of Tokyo RCAST, Meguro 153-0041, Tokyo, Japan}
\affil[b]{RIKEN Center for Advanced Intelligence Project, Chuo 103-0027, Tokyo, Japan}
\affil[c]{Sakurabashi Watanabe Hospital, Kita 530-0001, Osaka, Japan}

\begin{document} 
\maketitle

\section{DESCRIPTION OF PURPOSE}
\label{sec:intro}  

The automation of the diagnosis of coronary artery diseases is essential in order to fight what is one of the main causes of death worldwide. Such diagnosis begins in many cases with the detection of anomalies (e.g., plaques) in the coronary arteries from a coronary computed tomography angiography (CCTA).
Plaques are made up of cholesterol deposits and their buildup causes the inside of the arteries to narrow over time. Coronary plaques can be classified as being: non-calcified or calcified based on the amount of calcium in the lesion.
Unlike catheterization and intravascular methods, CCTA is non invasive, and allows capturing both types of calcified and non-calcified plaques. Non-calcified plaques are less stable and more likely to rupture, and thus, are associated with the risk of fatal cardiovascular events. However, while calcified plaques are easily recognizable, non-calcified plaques are difficult to estimate, as they have CCTA intensity values that can be mistaken with other elements present in the heart scan (e.g. fat, background tissues, artifacts).

Several methods for identifying plaques have been proposed in the past. Traditionally, a threshold intensity value was manually or adaptively selected in order to determine which voxels of the image potentially belong to plaque formations. However, these methods underperform in the case of non-calcified plaques, due to the ambiguous intensities in the image.
Recently, leveraging the power of annotated data, several deep learning-based methods were proposed, but their performance is still limited, as many non-calcified plaques are left undetected, and the shape of the plaques is too rough for a detailed analysis of the coronary lesion. More concretely, the shape of the artery wall and plaques is a decisive factor when assessing the patients' vulnerability to the lesion. We hypothesize that this is due to the presence of background noise and ambiguous intensity values in the periphery voxels of the artery wall that hinder the recognition of the actual plaques. In order to perform an accurate detection, a segmentation method should look for plaques within the artery wall region, so that peripheral fat and close-by vessels are ignored.
In a previous study, providing the inner and outer artery wall surfaces facilitated the \textit{manual} diagnosis of plaques in the coronary arteries. However, to the best of our knowledge, the effectiveness of this methodology has not been studied for the case of \textit{automatic} diagnosis, nor deep learning-based methods.

\begin{figure}[t]
\centering
\includegraphics[scale=0.4]{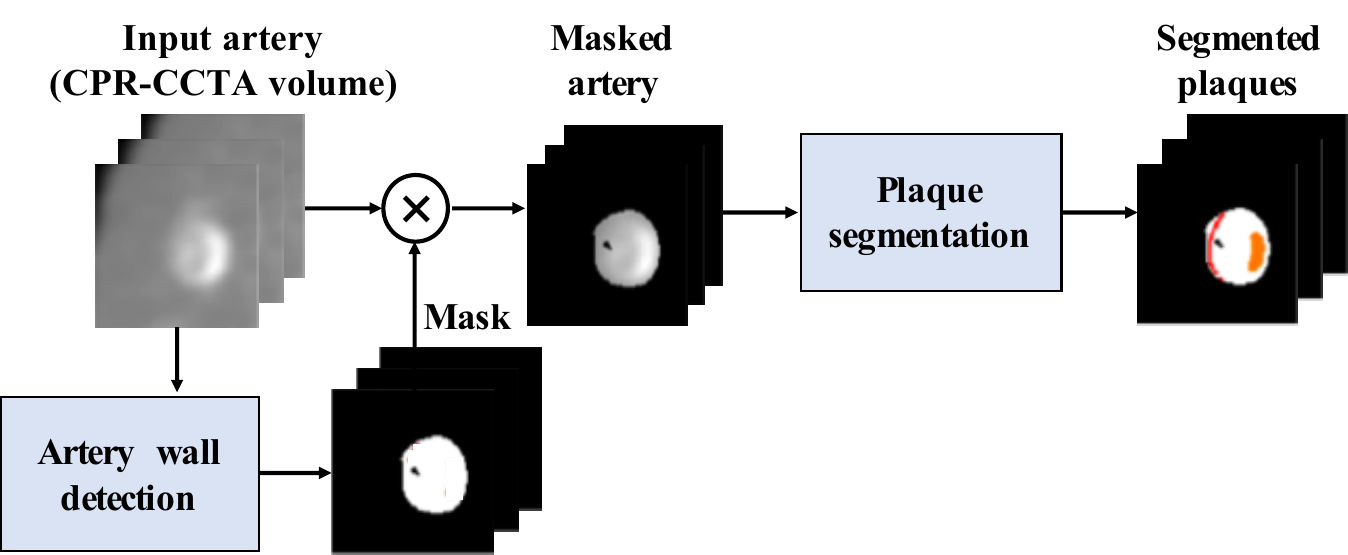}
\caption{Overview of our methodology. The artery wall is extracted prior to segmentation and used to create a mask. The mask removes voxels with ambiguous intensities facilitating the segmentation.}\label{fig:overview}
\vspace{-3mm}
\end{figure}

In this work, we propose a novel methodology for plaque segmentation in CPR-CCTA volumes (Fig.~\ref{fig:overview}). We mask the input voxels so only those within the artery wall are considered for segmentation. This mask can be obtained automatically via an inner-outer artery wall segmentation method.
Our contributions are three-fold:
\begin{itemize}
  \itemsep0em
  \item A novel methodology for plaque segmentation that masks the input CPR-CCTA voxels to boost accuracy.
  \item A deep learning-based implementation of our methodology with all the details for full-reproducibility.
  \item An exhaustive evaluation with different types of masks in order to validate our hypothesis, showing the potential of our findings.
\end{itemize}

\section{METHODOLOGY}

As shown in Figure~\ref{fig:overview}, our methodology consists of two main steps: extracting a mask of the artery wall, and segmenting the masked input to detect plaques. The input to our method are Curved Planar Reformation (CPR) volumes, which are extracted along the coronary artery paths of the CCTA scan. Anomalies such as plaques and stenosis are best observed in CPR scans, which map the artery walls following a straight line. Then, the output is a representation of the input in which each voxel has been assigned a label. Thus, in this study, we model the task of detecting plaques as a 4-class segmentation problem: \textit{calcified plaques}, \textit{non-calcified plaques}, \textit{artery wall}, and \textit{background} (\& \textit{lumen)}.

\subsection{Artery wall mask}

The first step is distinguishing the voxels located within the artery wall, from the remaining elements (background and lumen). As this paper is a study of the effectiveness of using an artery wall mask, we do not propose an artery wall detector per-se. Instead, we study the effect of employing masks of different accuracy range, being the most accurate mask the ground truth. Imperfect masks are created adding noise to the ground-truth mask (see Section~\ref{sec:experiments}). We also study the masks created with a state-of-the-art artery wall detector. While most artery extraction methods from CT scans focus solely on the inner wall, recent works allow extracting both inner and outer walls. HybridResUNet is a method for segmenting the inner and outer artery wall of an input CPR volume, which features a Weighted Hausdorff Distance loss and regularization for improving pixel connectivity.

Masks are created as follows. Given an input CCTA-CPR volume, with $S$ slices (cross-sections) of height $H$ and width $W$, the segmentation results of the artery wall segmentation is a set of voxels $V=\{v^s_{hw}|h\in[1,H],w\in[1,W],s\in[1,S]\}$. Voxels that belong to the inner or outer wall surfaces are denoted $I$ and $O$ respectively. Then, the mask is also a set of voxels $M=\{m^s_{hw}|h\in[1,H],w\in[1,W],s\in[1,S]\}$, and its values can be assigned slice by slice as:
\begin{equation}
    m^s_{hw}=
    \begin{cases}
        1, &\text{if $m^s_{hw}\in$ between $I^s$ and $O^s$;}\\
        0, &\text{otherwise.}
    \end{cases}
\end{equation}
where $I^s$ and $O^s$ represent the wall contours slice-wise. In our implementation, $H=W=512$ and $S$ varies by scan.

\subsection{Plaque segmentation}

To test our hypothesis, we adapted HybridResUNet for plaque segmentation as follows. First, we replaced the loss and regularization terms by the weighted \textit{Dice} loss. Then, to emphasize non-calcified plaques, the Dice loss for each class is weighted by the inverse of the pixel ratio of that class (the less common, the higher the weight).
We also modified the output channels from three to four when not using the ground-truth walls (background, artery wall, calcified plaque and non-calcified plaque). The decision of including a background class is not trivial; while perfect masks would only require predict the plaques within the artery wall, imperfect masks may let in voxels from either the lumen or the background, which need to be distinguished from the wall and plaque classes. Details on the data preprocessing and the training processing will be provided in the final manuscript.

\section{EXPERIMENTAL RESULTS}

\subsection{Dataset}

We evaluated our methodology with a non-public dataset of 100 CCTA scans, from where the CPR volumes for the LAD, LCX and RCA coronary arteries were extracted if available. Each artery is about $300\sim400$ slices long, which sums up to a total of 102983 slices in 291 arteries. The image acquisition details will be provided in the final manuscript.

Annotations for the voxels on each slice were provided by three expert radiologists. The annotation for a voxel $v$ with coordinates $ij$ in the $n^{th}$ slice of an artery is $v_{ij}^n \in \{0, 1, 2, 3, 4\}$, which represents the \textit{background}, \textit{lumen}, \textit{artery wall}, \textit{calcified plaque}, and \textit{non-calcified plaque} classes respectively.

We randomly selected one fourth of the scans (i.e., 25) as test data, in a way that the number of plaques is equally distributed.

\subsection{Experimental results}
\label{sec:experiments}

To examine our hypothesis that masking unrelated voxel values improves the plaque segmentation, we tried different masks configurations. While an ideal mask would leave the artery wall voxels exclusively, current wall detection methods have not reached that performance yet. Thus, we create imperfect masks by adding different levels of noise to the ground truth mask. Specifically, we randomly modify its position and thickness slice-wise in two noise configurations: small and large. Additionally, inner-outer artery walls predicted using HybridResUNet are also calculated. Including the case in which no mask is used, we consider a total of five different configurations.

Table~\ref{tab:quantitative} contains the Dice score of each class and the different mask configurations: (a) GT mask, (b) GT mask + small noise, (c) GT mask + large noise, (d) Wall segmentation (HybridResUNet), (e) No mask. Using a perfect mask (the ground-truth artery wall) produces highly accurate results, and does not require the use of a background class. On the next columns, adding noise affects accuracy, especially for the artery wall. 

\begin{table}[t]
\small
\caption{Class-wise Dice score comparison of different masking configurations in our methodology. The \textit{ground-truth wall} configuration does not use a background class.}\label{tab:quantitative}
\begin{center}
\begin{tabular}{l|c|c|c|c|c}
\hline
Class & (a) & (b) & (c) & (d) & (e) \\
\hline\hline
Background and lumen & N/A & 0.99 & 0.99 & 0.98 & 0.95 \\
Artery wall & 0.99 & 0.77 & 0.75 & 0.69 & 0.61 \\
Calcified plaque & 0.83 & 0.79 & 0.79 & 0.78 & 0.73 \\
Non-calcified plaque & 0.73 & 0.67 & 0.66 & 0.58 & 0.49 \\
\hline
\end{tabular}
\end{center}
\end{table}

\begin{figure}[t]
\centering
\includegraphics[scale=0.52]
{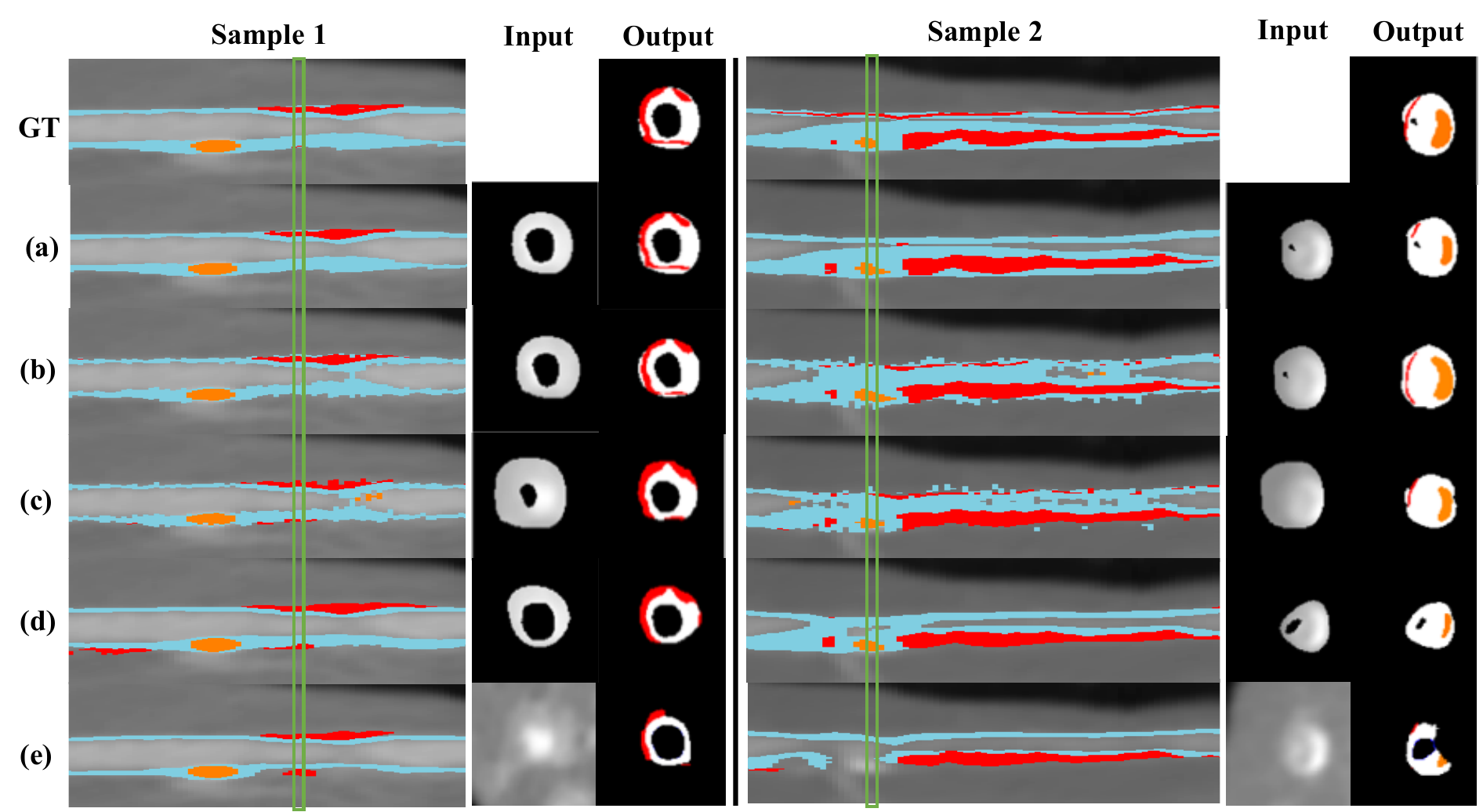}
\caption{Sample results for different mask configurations: (a) GT mask, (b) GT mask + small noise, (c) GT mask + large noise, (d) Wall segmentation (HybridResUNet), (e) No mask. The upper row shows the ground-truth (GT).}\label{fig:qualitative}
\end{figure}

Figure~\ref{fig:qualitative} shows a sample qualitative results of our experiments, repeated for the different mask configurations.
Each axial (cross-section) view corresponds to the area marked in green in its corresponding sagittal (side) view; one is the input target slice and next to it is our output (the first row is the ground-truth, so there is no input). Non-calcified plaques are colored in red, calcified plaques are orange, the artery wall is white, and the background class is black. In the side views, plaques and wall classes are overlaid to the normalized HU values (the wall is light blue for better visualization).
The sample shows both calcified and non-calcified plaques inside the artery wall. Applying the GT mask (a) yields a result highly similar to the ground-truth. While individual slices in (b) and (c) are not far from the ground-truth, the side views have a zigzag effect due to the randomness of the noise applied. However, despite the large noise in (c)'s input, the network remarkably obtains a shape closer to the ground-truth. In challenging samples, the mask extracted in (d) may differ from the ground-truth wall, providing inferior results than the other masked configurations. Finally, the unmasked segmentation case (e) provides the worst results, with incomplete artery walls and plaque protruding out the wall. The side views show an interesting phenomenon; inaccuracies in the mask may cause plaque false positives (FP). Specifically, the causes of non-calcified FPs is the fat tissue surrounding the arteries and partial volume effect. On the other hand, lumen voxels may cause calcified FPs when the mask is overly off center.

Our weakest masked configuration (d) provides results comparable to those of the related works, which are outperformed by the other configurations. A more detailed comparison will be provided in the final manuscript.



\subsection{Novel work to be presented}

To the best of our knowledge, this is the first work in plaque segmentation that examines in detail the performance boost obtained when masking the CCTA-CPR artery volumes. Almost voxel-perfect segmentations of calcified and non-calcified plaques can be obtained when using the accurate artery wall mask, which would allow for its practical use in a hospital. Moreover, the radiologists confirmed that our methodology was able to find existing plaques that were originally not included in the annotations.
On the contrary, highly inaccurate artery wall masks may produce false positive plaques in some cases, that should be removed via post-processing.

Apart from the resulting segmentations, these experiments also studied the learning process of the network. By limiting the network input to just the relevant voxels for segmentation, overfitting is prevented. The final manuscript will provide the learning graphs, as well as the histograms of the normalized HU values of the input for each configuration.

\section{CONCLUSIONS}

We proposed a novel methodology for plaque segmentation that ``masks'' the input CCTA-CPR artery volume to remove unrelated voxels and restrict the plaque detection to the artery wall. Our hypothesis that masking the input image improves segmentation is proven true in all the configurations, where most accurate masks provide better performance.

In this study, since both artery wall segmentation and plaque segmentation methods used in our experiments are deep neural networks, both methods could be trained together in an end-to-end manner. This way, the improvement in plaque segmentation would also be reflected in the wall segmentation. This will be addressed in our future work.

To the best of our knowledge, this is the first deep learning-based plaque segmentation study that explores the masking of the artery wall. The results obtained are promising and, as deep-learning based segmentation evolves, our findings will open the path for voxel-perfect plaque segmentation.

\textbf{Note: This work has not been published before.}




\end{document}